\documentclass[aps,pra,11pt]{revtex4-1}
\usepackage{mathtools}
\usepackage{units}
\usepackage{amssymb}
\usepackage{amsmath}
\usepackage{amsthm}
\usepackage{graphicx}

\newcommand{\pr}{\partial_r}
\newcommand{\pR}{\partial_R}
\newcommand{\pt}{\partial_t}

\newcommand{\px}{\partial_x}

\newcommand{\cpsi}{\bar{\psi}}
\newcommand{\cphi}{\bar{\phi}}
\newcommand{\cchi}{\bar{\chi}}

\newcommand{\expect}[1]{\left< #1 \right>}                                     
\newcommand{\braket}[2]{\left< #1 \vphantom{#2} \right| \left. #2 \vphantom{#1} \right>}
\newcommand{\matrixel}[3]{\left< #1 \vphantom{#2#3} \right| #2 \left| #3 \vphantom{#1#2} \right>}

\begin{document}
  
  \author{Axel Schild}
  \title{Time in quantum mechanics: \\ A fresh look at the continuity equation}
  \affiliation{ETH Z\"urich,  Laboratorium f\"ur Physikalische Chemie,  8093 Z\"urich, Switzerland}
  
  \begin{abstract}
    The local conservation of a physical quantity whose distribution changes 
    with time is mathematically described by the continuity equation.
    The corresponding time parameter, however, is defined with respect to an 
    idealized classical clock.
    We consider what happens when this classical time is replaced by a 
    non-relativistic quantum-mechanical description of the clock.
    From the clock-dependent Schr\"odinger equation (as analogue of the 
    time-dependent Schr\"odinger equation) we derive a continuity 
    equation, where, instead of a time-derivative, an operator occurs
    that depends on the flux (probability current) density of the clock.
    This clock-dependent continuity equation can be used to analyze the 
    dynamics of a quantum system and to study degrees of freedom that 
    may be used as internal clocks for an approximate description of the 
    dynamics of the remaining degrees of freedom.
    As an illustration, we study a simple model for coupled electron-nuclear 
    dynamics and interpret the nuclei as quantum clock for the electronic 
    motion.
    We find that whenever the Born-Oppenheimer approximation is valid, the 
    continuity equation shows that the nuclei are the only relevant clock for 
    the electrons.
  \end{abstract}
  
  \maketitle
  
  In many physical processes, the state of a system can be described by a 
  density distribution of a physical quantity which is locally 
  conserved, i.e., which is not created or destroyed during the process of 
  interest.
  The mathematical form of the conservation law for such a density is the 
  continuity equation \cite{aris1989}.
  It relates changes of the density of the conserved quantity to the divergence 
  of a vector field, called the flux density (or current density).
  The flux density represents the instantaneous motion of the density and, when 
  integrated over a surface, yields the flow of the density through that surface.
  As the continuity equation follows from the requirement of continuity alone \cite{aris1989},
  it is a very important relation for the mathematical description of nature.
  In non-relativistic quantum mechanics it holds for the probability density of 
  the particles \cite{boykin2000} and occurs for example in the hydrodynamic 
  formulation of quantum mechanics \cite{madelung1927,renziehausen2018}, in 
  time-dependent density functional theory \cite{runge1984}, and in the study of 
  nuclear and electron dynamics \cite{nafie2011,hermann2014,bredtmann2015,hermann2016}.
  Continuity is such a basic requirement that the continuity equation can also be used to 
  test theories, models, or numerical calculations for errors by comparing 
  the change of the density distribution to the expected fluxes.

  Although the continuity equation is a very general relation, it is typically 
  assumed that the dynamics is parametrized by a unique time, hence the 
  continuity equation is stated accordingly.
  However, the special status of time in quantum mechanics is currently 
  investigated \cite{page1983,briggs2000,braun2004,briggs2007,briggs2014,briggs2015,massar2015,boette2016,kitada2016,erker2017,malkiewicz2017} 
  and different approaches are developed that view time mostly as an emergent 
  property, assuming that quantum mechanics is fundamentally timeless.
  One of these is the Page-Wootters approach \cite{page1983,moreva2014,moreva2015,giovannetti2015,moreva2017,marletto2017,bryan2018}
  where the timeless universe (a closed system containing all relevant degrees 
  of freedom) is partitioned into a clock and a system of interest.
  This system has to be entangled with the clock, and there has to exist a good 
  clock in the sense that it has many distinguishable states but little 
  interaction with the system \cite{marletto2017}.
  The concept of time can then be found as a (classical) conditional variable.
  
  In this article, we use a different but related approach which allows us 
  to include quantum-mechanical effects of the clock:
  In the Briggs-Rost approach \cite{briggs2000,briggs2015}, a closed composite 
  (i.e., consisting of at least two degrees of freedom) universe is also 
  separated into a clock and a system which depends conditionally on (and which 
  is entangled with) the clock.
  Then,  from the time-independent Schr\"odinger equation (TISE) of the system 
  together with the clock, the time-dependent Schr\"odinger equation (TDSE) of
  the system can be obtained in the classical limit for the clock.
  It follows that the TDSE is a quantum-classical equation.
  Consequently, the time-dependent continuity equation is a quantum-classical 
  relation, too, and we may ask if a fully quantum-mechanical equivalent can be 
  found.
  
  To investigate this question, we use the Briggs-Rost approach in the 
  language of the Exact Factorization \cite{abedi2010,briggs2015}, where the 
  joint probability density for system and clock is separated into a marginal 
  probability density for the clock and a conditional probability density for 
  the system, which conditionally depends on the state of the clock.
  Thus, time has a similar status as it has in the Page-Wootters approach.
  However, no assumptions or constructions are necessary and all results can 
  directly be derived from the time-independent equation of motion, which in
  our case is the TISE.
  This will allow us to find a fully quantum-mechanical continuity equation by 
  replacing the TDSE of the system with a clock-dependent Schr\"odinger equation 
  (CDSE), which becomes a TDSE in the classical limit of the clock wavefunction.
  Instead of a conditional time parameter, both the CDSE and its continuity 
  equation depend on the (quantum) state of the clock.
  
  Such a quantum-mechanical continuity equation is interesting from a 
  fundamental point of view, because it illustrates the quantum-mechanical 
  nature of the clock that is used to track the dynamics of the system.
  Our focus in this article, however, is to illustrate a possible practical
  purpose of the quantum-mechanical continuity equation:
  It can be used to analyze a dynamics and to find degrees of freedom that 
  can be used as clocks for other degrees of freedom, possibly allowing to find 
  approximate simulation methods in the spirit of the Born-Oppenheimer 
  approximation.
  Born-Oppenheimer-type approaches are used in many different ways, e.g.\ for 
  quantizing constrained systems \cite{kaplan1997}, for Rydberg states 
  \cite{remacle1998}, for quantum heat transfer \cite{wu2011}, for a model of 
  atoms in a oscillator and lattice trap \cite{sorensen2012}, or for the 
  dynamics of H$_2$ in carbon nanotubes \cite{martell2017}.
  The clock-dependent continuity equation shows, in a novel way, how the 
  approximation works:
  As an example, we consider a simple model of the coupled electron-nuclear 
  dynamics during a proton-coupled electron transfer process and treat the 
  nuclei as a quantum clock for the electronic motion.
  In this case, there are two clocks for the electrons: 
  An external clock (to which also the nuclear motion is referred) and an 
  internal clock given by the nuclear wavefunction.
  In the continuity equation, the change of the electron density w.r.t.\ these 
  two clocks shows clearly under which conditions only the internal clock is 
  relevant for the  electronic motion. 
  These conditions corresponds to situations where the Born-Oppenheimer 
  approximation is applicable, i.e., where the change of the nuclear 
  configuration is enough to represent the electron dynamics.
  Of course, the conditions for which the Born-Oppenheimer approximation works
  are well known, but our analysis in terms of different clocks gives both a new 
  point of view on this familiar method and it allows a generalization of the 
  approach to other problems where similar separations of time scales may be 
  helpful.
  As a by-product, treating the nuclei as a clock for the electrons also 
  sheds a new light on the riddle of the vanishing electronic flux density in 
  the Born-Oppenheimer approximation \cite{barth2009}, as explained below.
  
  {\bf The time-dependent continuity equation}
  
  Before developing the idea of time measured by means of a quantum clock,
  we first review the time-dependent continuity equation in quantum mechanics.
  Let $\rho(x|t) \in \mathbb{R}$ be the distribution of a conserved 
  continuous physical quantity depending on spatial coordinates 
  $x \in \mathbb{R}^3$ and on time $t$.
  The change of $\rho(x|t)$ in some compact volume $\Omega$ has to correspond 
  to the flux through the surface $\partial \Omega$ of the volume.
  This requirement is stated as \cite{aris1989}
  \begin{align}
    \pt \int_{\Omega} \rho(x|t) \, dx + \int_{\partial \Omega} j(x|t) \cdot dS = 0,
    \label{eq:eoc_gen_int}
  \end{align}
  where the vector field $j(x|t) \in \mathbb{R}^3$ is the flux density (or 
  current density) and where $\pt$ is the derivative w.r.t.\ time.
  By means of Stoke's theorem and the requirement that \eqref{eq:eoc_gen_int}
  is true for any volume, we obtain the continuity equation \cite{aris1989}
  \begin{align}
    \pt \rho(x|t) + \px \cdot j(x|t) = 0
    \label{eq:eoc_gen_dif}
  \end{align}
  with the derivative vector w.r.t.\ the components of $x$ denoted as $\px$.
  Eq.\ \eqref{eq:eoc_gen_dif} relates changes of $\rho(x|t)$ in time to the 
  divergence of $j(x|t)$.
  If the state $\psi(x|t)$ of a particle is described by the TDSE
  \begin{align}
    i \hbar \pt \psi(x|t) = \left( \frac{\hbar^2}{2m} (-i \px + A(x|t))^2 + V(x|t) \right) \psi(x|t)
    \label{eq:tdse0}
  \end{align}
  with scalar potential $V \in \mathbb{R}$ and vector potential 
  $A  \in \mathbb{R}^3$, and if $|\psi(x|t)|^2 = \rho(x|t)$ is the probability 
  density of this particle, the probability flux density $j(x|t)$ in the 
  continuity equation \eqref{eq:eoc_gen_dif} is identified with
  \begin{align}
    j(x|t) = \frac{\hbar}{m} \left( \operatorname{Im}\left( \cpsi(x|t) \px \psi(x|t) \right) + A(x|t) |\psi(x|t)|^2 \right),
    \label{eq:ufd}
  \end{align}
  where $\cpsi$ denotes the complex-conjugate of $\psi$.
  
  We note that the flux density $j(x|t)$ is defined via \eqref{eq:eoc_gen_dif} 
  only up to the addition of a vector field $j_{\perp}(x|t)$ for which 
  $\px \cdot j_{\perp}(x|t) = 0$.
  Hence, \eqref{eq:ufd} is not a unique definition if only the continuity 
  equation and the TDSE are known \cite{hodge2014}.
  It is, however, the position representation of the quantum-mechanical 
  operator corresponding to the classical flux density, and the continuity
  equation may be derived from this operator \cite{boykin2000}.
  
  {\bf The clock-dependent Schr\"odinger equation}
  
  To describe the system and its clock quantum mechanically, we follow the 
  developments presented in \cite{briggs2000,briggs2015}.
  In these articles, it was shown that the TDSE \eqref{eq:tdse0} can be obtained 
  from the TISE for the considered system together with the clock that is used 
  to measure the time parameter.
  Two requirements were necessary:
  The energy of the clock has to be much larger than that of the system so that
  its state is negligibly disturbed by the system, and the classical limit
  has to be taken for the wavefunction of the clock.
  Only in this case a time parameter can be defined which corresponds to a 
  ``universal'' reference time.
  For our discussion, we do not need to make any of these assumptions, but we 
  keep the clock fully quantum-mechanical.
  Our derivation is based on the Exact Factorization method 
  \cite{abedi2010,abedi2012} that was also used in \cite{briggs2015}, but we 
  explicitly take into account the gauge freedom that appears in this theory.
  
  We start from the TISE
  \begin{align}
    \left(-\frac{\hbar^2 \partial_R^2}{2 M} + H_{\rm S} \right) \psi(R,r) = E \psi(R,r)
    \label{eq:tise}
  \end{align}
  for a closed system (the ``universe'') with energy $E$.
  The Hamiltonian of the system is given as
  \begin{align}
    H_{\rm S} = -\frac{\hbar^2 \partial_r^2}{2 m} + V(R,r),
  \end{align}
  where $R, r \in \mathbb{R}^3$ are the coordinates of two particles and where 
  the gradients w.r.t.\ $R$ and $r$ are denoted as $\partial_R$ and $\partial_r$, 
  respectively.
  The restriction of the notation to two particles leads to a simplification of 
  the equations.
  The generalization to many particles is straightforward and is discussed 
  briefly below.
  The wavefunction $\psi$ shall be normalized according to 
  \begin{align}
    \braket{\psi(R,r)}{\psi(R,r)} = 1, \label{eq:norm}
  \end{align}
  where $\braket{\cdot}{\cdot}$ denotes the scalar product w.r.t.\ coordinates 
  $R$ and $r$.
  In the Exact Factorization ansatz \cite{abedi2010,gidopoulos2014} the 
  wavefunction is written as a product
  \begin{align}
    \psi(R,r) = \chi(R) \phi(r|R) \label{eq:psi}
  \end{align}
  of two amplitudes $\chi(R)$ and $\phi(r|R)$, which fulfill the following 
  properties:
  The function 
  \begin{align}
    |\chi(R)|^2 := \braket{\psi}{\psi}_r
  \end{align}
  is the marginal probability density which represents the probability of 
  finding a particle of mass $M$ at $R$ independent of where the particle with 
  mass $m$ is.
  The symbol $\braket{\cdot}{\cdot}_r$ indicates the scalar product w.r.t.\ the 
  coordinate $r$ only.
  Also,
  \begin{align}
    \phi(r|R) := \psi(R,r) / \chi(R)
  \end{align}
  yields the  conditional probability density $|\phi(r|R)|^2$ of finding a 
  particle with mass $m$ at $r$, given there is a particle with mass $M$ at $R$.
  It obeys the partial normalization condition
  \begin{align}
    \braket{\phi(r|R)}{\phi(r|R)}_r \stackrel{!}{=} 1. \label{eq:pnc}
  \end{align}
  This condition has to be valid all values of $R$.
  
  The function $\chi(R)$ is interpreted as the wavefunction of the clock,
  whereas the function $\phi(r|R)$ is interpreted as the wavefunction of the 
  system which depends conditionally on the configuration $R$ of the clock.
  The equation of motion for the clock is then \cite{abedi2010,gidopoulos2014}
  \begin{align}
    \left( \frac{\hat{P}^2}{2 M} + \epsilon(R) \right) \chi = E \chi
    \label{eq:chi}
  \end{align}
  and the equation of motion for the system is 
  \begin{align}
    \hat{C} \phi = \left(\hat{H}_{\rm S} + \hat{U} - \epsilon(R) \right) \phi.
    \label{eq:phi}
  \end{align}
  In the equation for the system the clock-dependent operator
  \begin{align}
    \hat{C} = \frac{1}{M} \frac{\hat{P} \chi}{\chi} \cdot \hat{P}^{\dagger} \label{eq:C}
  \end{align}
  and the kinetic operator
  \begin{align}
    \hat{U} = \frac{ ( \hat{P}^{\dagger} )^2}{2 M} \label{eq:U}
  \end{align}
  occur which both operate on the conditional variable $R$, because the 
  momentum operator $\hat{P}$ is defined as
  \begin{align}
    \hat{P} := \hbar \left( -i \partial_R + A \right) \label{eq:P}
  \end{align}
  with hermitian adjoint
  \begin{align}
    \hat{P}^{\dagger} = \hbar \left( i \partial_R + A \right) . \label{eq:Pd}
  \end{align}
  The real-valued scalar potential $\epsilon(R) \in \mathbb{R}$ is obtained from these operators as
  \begin{align}
    \epsilon(R) := \matrixel{\phi}{\hat{H}_{\rm S} + \hat{U} - \hat{C} }{\phi}_r, \label{eq:scapot}
  \end{align}
  whereas the real-valued vector potential $A(R) \in \mathbb{R}^3$ is given by
  \begin{align}
    A(R) := -i \braket{\phi}{\partial_R \phi}_r. \label{eq:vecpot}
  \end{align}
  
  The operators $\hat{P}$ and $\hat{P}^{\dagger}$ are hermitian adjoints for 
  $\chi$ w.r.t.\ $R$-space, because
  \begin{align}
    \braket{\hat{P} \chi}{\chi}_R \equiv \int \chi \hat{P}^{\dagger} \cchi dR = \braket{\chi}{\hat{P} \chi}_R,
  \end{align}
  but also for $\phi$ w.r.t.\ $r$-space:
  If we interpret the definition \eqref{eq:vecpot} of the vector potential $A$
  such that $A$ is a functional of $\phi$, we find from the partial 
  normalization condition \eqref{eq:pnc} that $A[\phi] = -A[\cphi]$.
  It follows that
  \begin{align}
    \braket{\hat{P} \phi}{\phi}_r = \braket{\phi}{\hat{P} \phi}_r = 2 \hbar A[\phi].
  \end{align}
  We note that if the scalar potential \eqref{eq:scapot} is taken as a 
  functional of $\phi$, we have $\epsilon[\phi] = \epsilon[\cphi]$.
  
  From the ansatz \eqref{eq:psi}, $\chi(R)$ and $\phi(r|R)$ are defined up to a phase 
  factor $\theta(R)$.
  Choosing $\theta(R)$ means choosing a gauge, as the equations of motion 
  \eqref{eq:chi}, \eqref{eq:phi} are unchanged if the marginal amplitude, the 
  conditional amplitude, and the vector potential are replaced by
  \begin{align}
    \chi'(R)   &= \chi(R)   e^{-i \theta(R)} \nonumber \\
    \phi'(r|R) &= \phi(r|R) e^{ i \theta(R)} \nonumber \\
    A'(R)      &= A(R) + \partial_R \theta(R).
    \label{eq:gauge}
  \end{align}
  
  For a given state $\chi(R)$ of the clock, \eqref{eq:phi} yields the state of 
  the system $\phi$.
  If we take 
  \begin{align}
    \chi(R) =: e^{-i W(R)} =: |\chi(R)| e^{-i S(R)}
  \end{align}
  with $W(R) \in \mathbb{C}$ and $S(R) \in \mathbb{R}$, we can 
  rewrite \eqref{eq:phi} either as
  \begin{align}
    i \hbar \hat{c}_W \phi = \left( \hat{H}_{\rm S} + \hat{U} - \epsilon(R) \right) \phi \label{eq:cdse_w}
  \end{align}
  with
  \begin{align}
    \hat{c}_W := \frac{\hbar}{M} (-\pR W + A) (\pR - iA) \label{eq:op_cw},
  \end{align}
  or as 
  \begin{align}
    i \hbar \hat{c}_S \phi = \left( \hat{H}_{\rm S} + \hat{U} + \hat{u}_S - \epsilon(R) \right) \phi \label{eq:cdse_s}
  \end{align}
  with
  \begin{align}
    \hat{c}_S &:= \frac{\hbar}{M} (-\pR S + A) (\pR - iA) \label{eq:op_cs} \\
    \hat{u}_S &:= \frac{\hbar^2}{2 M} \frac{\pR |\chi|^2}{|\chi|^2} (i A - \pR).
  \end{align}
  We note that the above equations can easily be extended to many particles by 
  redefining the coordinates as $R \in \mathbb{R}^{3 n_{\rm C}}$, 
  $r \in \mathbb{R}^{3 n_{\rm S}}$ if the clock and system consist of 
  $n_{\rm C}$ and $n_{\rm S}$ particles, respectively.
  The above equations can directly be used if the number of components of the 
  vector quantities is adjusted and if the coordinates are mass-scaled such that 
  only two masses $M$ and $m$ occur.
  Alternatively, individual vector potentials and flux densities for each 
  particle can be defined.
  
  We call both \eqref{eq:cdse_w} and \eqref{eq:cdse_s} the clock-dependent 
  Schr\"odinger equation (CDSE) in analogy to the TDSE, because they become
  TDSEs if the classical limit for the clock is taken.
  How to take this classical limit is discussed in \cite{briggs2015}, and we 
  only sketch here the steps that need to be done, using some results of an
  analysis of the adiabatic limit in the Exact Factorization \cite{eich2016}.
  First, we introduce a parameter $\mu := \sqrt{m/M}$ which is the ratio of the 
  mass of the system to that of the clock.
  If the clock is heavy compared to the system ($\mu$ is small), the 
  quantization of the clock becomes negligible and the clock behaves 
  classically.
  We can then expand $W(R)$ in powers of $\mu$,
  \begin{align}
    W(R) = \frac{1}{\mu} \sum\limits_{n=0}^{\infty} \mu^n W_n.
  \end{align}
  To lowest order in $\mu$, it can be shown \cite{briggs2015,eich2016} that 
  the equation of motion for the clock \eqref{eq:chi} becomes a classical 
  time-independent Hamilton-Jacobi equation and $W_0$ is the classical 
  real-valued action.
  Next, we consider the CDSE \eqref{eq:cdse_w} and use the result of 
  \cite{eich2016} that $\hat{U}$ is of order $\mu^2$ and can be neglected, 
  while $\hat{c}_W$ is of order $\mu$ and is kept.
  By choosing the gauge where the vector potential vanishes, $A = 0$, the CDSE 
  becomes
  \begin{align}
    i \hbar  \hat{c}_W \phi \rightarrow i \hbar \frac{P}{M} \pR \phi \approx \left( \hat{H}_S - \epsilon(R) \right) \phi,
    \label{eq:naja}
  \end{align}
  where $P$ is the classical momentum of the clock defined via the derivative 
  $\pR W_0$ of the classical action of the clock.
  We can now define a time $t$ that parametrized the position $R$ of 
  the clock, such that the classical momentum is $P = M \pt R(t)$, and
  \begin{align}
    i \hbar \frac{P}{M} \pR \phi = i \hbar \partial_t \phi.
  \end{align}
  Thus, \eqref{eq:naja} is a TDSE.
  
  This sketch of the derivation of the TDSE from the CDSE leaves out many 
  technical details, e.g.\ the change of the scalar potential $\epsilon(R)$
  when the classical limit is taken or how to perform the classical limit in 
  a gauge invariant way, i.e., including the vector potential $A(R)$.
  While some of those details are explained in \cite{briggs2015,eich2016},
  some are still open problems  and may be rewarding topics for future research.
  We need not be concerned with them here, however, because we keep the clock
  fully quantum mechanical and derive the continuity equation for such a 
  quantum clock.
  
  The operators $\hat{c}_W$ and $\hat{c}_S$ have some interesting properties:
  They are gauge invariant, as a change of gauge \eqref{eq:gauge} yields
  \begin{align}
    \hat{c}_{S/W}' \phi'(r|R) = e^{i \theta(R)} \hat{c}_{S/W} \phi(r|R),
  \end{align}
  and $i \hbar \hat{c}_{S}$ is (like $i \hbar \pt$) hermitian for $\phi$
  w.r.t.\ the $r$-space,
  \begin{align}
    \braket{\phi}{i \hbar \hat{c}_{S} \phi}_r = \braket{i \hbar \hat{c}_{S} \phi}{\phi}_r,
  \end{align}
  because of the normalization condition $\eqref{eq:pnc}$ that $\phi$ has to 
  fulfill.
  However, in contrast to $\pt$ the operators $\hat{c}_{W/S}$ are in general 
  complex.
  
  {\bf Time-reversal invariance}
  
  Before using the CDSE to derive a continuity equation, we would like to 
  comment on time-reversal invariance and how it occurs in the CDSE.
  For the TDSE, it is well-known \cite{messiah1999} that it is equivalent to 
  either solve the TDSE with time $t$, or to solve its complex-conjugate 
  equation with time $-t$, provided the sign of a possible vector potential 
  occurring in the canonical momentum operator is changed.
  Hence, solving the TDSE \eqref{eq:tdse0} or solving
  \begin{align}
    -i \hbar \partial_{-t} \bar{\psi}(x,-t) = \left( \frac{\hbar^2}{2m} (i \px - A(x,-t))^2 \right) \bar{\psi}(x,-t)
  \end{align}
  for the same initial condition $\psi(x,t_0)$ yields the same probability 
  density $|\psi|^2$ and, except for a change of sign, the same probability 
  flux density \eqref{eq:ufd}.
  
  From the clock-dependent point of view, time-reversal invariance originates 
  from invariance of the TISE of the universe, \eqref{eq:tise}, w.r.t.\ to 
  complex conjugation.
  Taking, for example, the complex-conjugate of the CDSE \eqref{eq:cdse_s}
  yields an equation equivalent to \eqref{eq:cdse_s}, provided one realizes 
  that $A[\phi]$ has to be replaced with $A[\cphi] = -A[\phi]$, and provided 
  one changes the sign of the clock's phase $S(R)$, making it run in the 
  ``reverse'' direction.
  Is this sense the CDSE is clock-reversal invariant, but this is a rather 
  trivial consequence of the requirement that the universe is in a static
  state.
  
  {\bf The clock-dependent continuity equation}
  
  To obtain the fully quantum-mechanical continuity equation for the density 
  $|\phi(r|R)|^2$ of the system which depends on the system coordinates $r$, on 
  the clock coordinates $R$, and also on the state of the clock $\chi$, we work
  with $\hat{c}_S$ and treat it as analogue of the time-derivative operator.
  Its action on the conditional density $|\phi(r|R)|^2$ can be defined as
  \begin{align}
    \cphi \hat{c}_S \phi + \phi \hat{c}_S^{\dagger} \cphi
      =: \hat{c} |\phi|^2 \label{eq:c}
  \end{align}
  with purely real operator 
  \begin{align}
    \hat{c} = \frac{\hbar}{M} (-\pR S + A) \pR.
  \end{align}
  From the CDSE for $\phi$ and for its complex-conjugate $\cphi$, we find
  \begin{align}
    \hat{c} |\phi|^2 
    + \pr \cdot j[\phi](r|R) 
    + \left( \pR + \frac{\pR |\chi|^2}{|\chi|^2} \right) \cdot J[\phi,-A](r|R) 
    = 0
    \label{eq:ccc}
  \end{align}
  with the flux densities $j[\phi]$ and $J[\phi,-A]$ being of the usual form 
  of \eqref{eq:ufd}, given for a general function $f \in \mathbb{C}$ as
  \begin{align}
    j[f]   &= \frac{\hbar}{m} \operatorname{Im}\left( \bar{f} \pr f \right) \\
    J[f,\pm A] &= \frac{\hbar}{M} \left( \operatorname{Im}\left( \bar{f} \pR f \right) \pm A |f|^2 \right).
  \end{align}
  Here, $j[\phi]$ appears because of the Hamiltonian $\hat{H}_{\rm S}$ of the system
  alone and is the flux density of the system w.r.t.\ the system variables, 
  while $J[\phi,-A]$ appears because of the operator $\hat{U}$ and is the flux 
  density of the system w.r.t.\ the clock variables.
  We note that $J[f,\pm A]$ (as well as $j[f]$) are invariant w.r.t.\ the choice 
  of gauge \eqref{eq:gauge}.
  With these definitions, we can also write the operator $\hat{c}$ as
  \begin{align}
    \hat{c} = \frac{1}{|\chi|^2} J[\chi,+A] \cdot \pR.
  \end{align}
  
  Eq.\ \eqref{eq:ccc} is a clock-dependent continuity equation in the space of 
  the coordinates of the system which, however, depends also conditionally on 
  the coordinates of the clock.
  Three flux densities occur in this equation:
  the flux density of the system w.r.t.\ its coordinates, the flux density of 
  the system w.r.t.\ the clock coordinates, and, as part of the $\hat{c}$-operator, 
  the flux density of the clock.
  The $\hat{c}$-operator yields the change of the density $|\phi|^2$ w.r.t.\ the 
  clock variables, weighted by the velocity components (flux density divided by 
  density) in the respective direction.
  What is also notable about \eqref{eq:ccc} is that there is a modification of 
  the divergence of $J[\phi,-A]$ by the additional term $(\pR |\chi|^2)/|\chi|^2$.
  This could be avoided if we had used the complex-valued 
  function $W(R)$ instead of the real-valued functions $S(R)$ and 
  $|\chi(R)|$, i.e., if we had used $\hat{c}_W$ instead of $\hat{c}_S$.
  The disadvantage of this choice would have been that we could not have written
  \eqref{eq:c} with a real-valued operator $\hat{c}$ acting on $|\phi|^2$.
  
  To obtain a continuity equation for the system alone, we need to average over 
  the degrees of freedom of the clock.
  This is done by multiplication of \eqref{eq:ccc} with the probability 
  distribution of the clock $|\chi|^2$ and subsequent integration 
  over the clock configurations $R$.
  Due to the modified divergence w.r.t.\ the clock coordinates, the contribution 
  of the flux density $J[\phi,-A]$ cancels,
  \begin{align}
    \int |\chi|^2 \left( \pR + \frac{\pR |\chi|^2}{|\chi|^2} \right) \cdot J[\phi,-A](r|R) dR = 0,
  \end{align}
  and we obtain 
  \begin{align}
    \dot{\rho}_S[\chi](r) + \pr \cdot j_S[\chi](r) = 0 \label{eq:cc}
  \end{align}
  with the clock-averaged change of the density defined as
  \begin{align}
    \dot{\rho}_S[\chi](r) := \int |\chi(R)|^2 \hat{c} |\phi(r|R)|^2 dR
  \end{align}
  and with the clock-averaged flux density
  \begin{align}
    j_S[\chi](r) := \int |\chi(R)|^2 j[\phi](r|R) dR.
  \end{align}
  
  The clock-dependent continuity equations \eqref{eq:ccc} and \eqref{eq:cc} 
  are the main results of this article.
  The equation for the system alone, \eqref{eq:cc}, is especially interesting,
  as it is the direct analogue of the time-dependent continuity equation.
  Instead of the usual time-dependence, there is a functional dependence on 
  the state of the clock $\chi$ in both the change of the density $\dot{\rho}_S$ 
  and in the flux density $j_s$ which takes into account the quantum-mechanical 
  nature of the clock.
  
  The clock-dependent continuity equation \eqref{eq:cc} can be useful to 
  analyze a quantum dynamics, as illustrated in the next section.
  However, as we started from a stationary state of the universe, it is not 
  obvious from \eqref{eq:ccc} and \eqref{eq:cc} how to obtain a non-trivial 
  dynamics of the system.
  Here, non-trivial means that the terms of the continuity equations are not
  individually zero.
  If the universe (which is assumed to be a closed system) is in an eigenstate 
  of zero total angular momentum, its wavefunction can be chosen to be real and 
  hence both the clock wavefunction $\chi$ and the system wavefunction $\phi$ 
  may be chosen to be real.
  Then, the (gauge-invariant) flux densities $j[\phi]$, $J[\phi,-A]$, and 
  $J[\chi,+A]$ in \eqref{eq:ccc} vanish and there is no dynamics. 
  However, there are cases where there always exists a non-zero vector potential
  leading to non-zero flux densities, even for eigenstates \cite{requist2016}.
  This is the case if the universe is in a state of non-zero angular momentum, 
  and then it may be possible to obtain a non-trivial dynamics. 
  We tested a model of a universe having only harmonic interactions, finding
  that for an eigenstate of the universe with non-zero integer angular 
  momentum, a gauge $A=0$ is in general not possible.
  Nevertheless, after averaging over the clock wavefunction, we find that the 
  terms in \eqref{eq:cc} are still individually zero.
  This test is of course not conclusive and further investigations are 
  necessary.
  
  A clock-dependent measurement has, however, not only a system and a clock 
  that are involved, but also an observer which measures both the state of the 
  clock and of the system. 
  Taking the act of measurement into account may lead to a non-trivial dynamics 
  even if the universe is in an eigenstate, and hence may change the 
  interpretation of the clock-dependent continuity equations.
  It may thus be necessary to take the view of an internal observer that is 
  measuring the state of the clock and of the system, and some ideas of how to 
  obtain such a consistent timeless theory with evolution exist
  \cite{malkiewicz2017}.
  
  In the following example application we avoid the question of the origin of 
  the dynamics from a universe-internal point of view altogether by assuming 
  that the dynamics is measured w.r.t.\ some external classical time (i.e., the 
  dynamics is the solution of a TDSE).
  This is the typical experimental situation, but it is understood that the lack
  of a fully internal description is unsatisfying from a fundamental point of 
  view and calls for further research.
  
  {\bf Application}
  
  To understand the clock-dependent continuity equation better, we consider 
  as an example a molecular dynamics, i.e., a dynamics of nuclei and electrons, 
  which is generated by an external interaction (that need not be present 
  anymore) and which can be described by a TDSE.
  Our aim is to study the electron dynamics with respect to the external 
  classical time and with respect to the nuclei, which are treated as an 
  internal clock.
  As the ansatz is similar to the Born-Oppenheimer approach to molecular 
  dynamics, the example allows us to understand better how the Born-Oppenheimer 
  approximation works and shows how the idea of separating ``slow'' and ``fast''
  degrees of freedom can be transfered to other problems.
  Additionally, it gives new insights into the puzzling problem that the 
  electronic continuity equation seems to be violated in the limit of the 
  Born-Oppenheimer approximation \cite{barth2009}.
  
  The state of the molecule is described by a wavefunction $\psi(R,r|t)$
  depending on nuclear coordinates $R$, on electronic coordinates $r$, and on an
  external time $t$, i.e., it is assumed that the considered dynamics is 
  governed by the TDSE
  \begin{align}
    i \hbar \pt \psi = \left(-\frac{\hbar^2 \pR}{2M} + \hat{H}_{\rm el} \right) \psi \label{eq:tdse}
  \end{align}
  with electronic Hamiltonian $\hat{H}_{\rm el}$ (corresponding to the system Hamiltonian 
  $\hat{H}_{\rm S}$, cf.\ \eqref{eq:tise}) containing the electronic kinetic energy operator 
  and the scalar interaction potential.
  Initialization of the dynamics and measurement of $t$ can happen by means of a 
  suitable external interaction, e.g.\ a strong ultrashort laser pulses
  that can act as a good clock \cite{braun2004}.
  
  In analogy to \eqref{eq:psi}, we make the factorization ansatz
  \begin{align}
    \psi(R,r|t) = \chi(R|t) \phi(r|R,t)
  \end{align}
  with partial normalization condition $\braket{\phi}{\phi}_r \stackrel{!}{=} 1
  \forall R, t$, where the equations of motion are \cite{abedi2010}
  \begin{align}
    i \hbar \pt \chi &= \left( -\frac{\hbar^2}{2M} \left( i \pR + A(R|t) \right)^2 + \epsilon(R|t) \right) \chi \label{eq:chi_t} \\
    \left( i \hbar \pt + \hat{C} \right) \phi &= \left( \hat{H}_{\rm el} + \hat{U} - \epsilon(R|t) \right) \phi \label{eq:phi_t}
  \end{align}
  with time-dependent vector potential $A(R|t)$ as defined in \eqref{eq:vecpot},
  with the operators $\hat{C}$, $\hat{U}$ as defined in \eqref{eq:C}, 
  \eqref{eq:U}, and with time-dependent scalar potential
  \begin{align}
    \epsilon(R|t) = \matrixel{\phi}{\hat{H}_{\rm el} + \hat{U} - \hat{C} - i \hbar \pt}{\phi}_r.
  \end{align}
  For a change of gauge \eqref{eq:gauge}, the time-dependent scalar potential
  transforms as $\epsilon'(R|t) = \epsilon(R|t) + \pt \theta(R,t)$.
  The continuity equation following from the TDSE for $\psi$, integrated over 
  $R$, gives 
  \begin{align}
    0 &= \expect{|\chi|^2 \pt |\phi|^2}_R + \expect{|\phi|^2 \pt |\chi|^2}_R
        + \expect{ |\chi|^2 \pr \cdot j[\phi]}_R. \label{eq:ce8}
  \end{align}
  Also, \eqref{eq:chi_t} is a normal TDSE for $\chi(R|t)$, hence the 
  corresponding continuity equation is 
  \begin{align}
    0 = \pt |\chi|^2 + \pR \cdot J[\chi,A]. \label{eq:ce9}
  \end{align}
  Inserting \eqref{eq:ce9} for $\pt |\chi|^2$ in \eqref{eq:ce8} yields the 
  $t$- and clock-dependent continuity equation
  \begin{align}
    0 &= \expect{|\chi|^2 \pt |\phi|^2}_R + \expect{J[\chi,A] \cdot \pR |\phi|^2}_R + \pr \cdot \expect{|\chi|^2 j[\phi]}_R
    \label{eq:ccc_t}
  \end{align}
  where a partial integration was used once for the $J$-dependent term.
  $\expect{|\chi|^2 j[\phi]}_R$  is the electronic flux density, and a 
  comparison with \eqref{eq:cc} shows that $\expect{|\chi|^2 \pt |\phi|^2}_R$ 
  can be interpreted as the (averaged) change of the electronic density w.r.t.\
  the external time, and $\expect{J[\chi,+A] \cdot \pR |\phi|^2}_R$ is the 
  change of the electronic density w.r.t.\ the internal clock.
  
\begin{figure}
    \centering
    \includegraphics[width=0.8\textwidth]{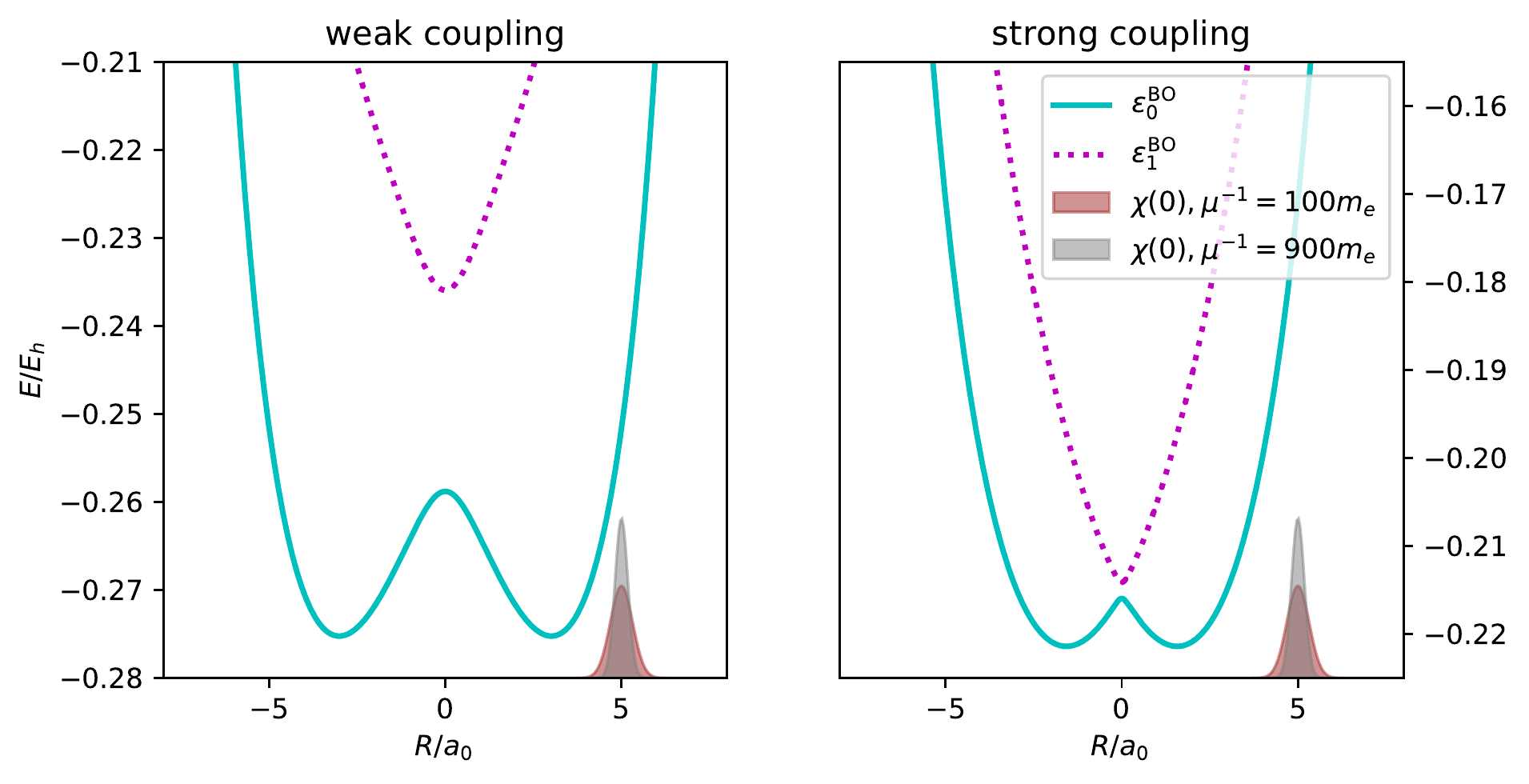}
    \caption{The two lowest Born-Oppenheimer potential energy surfaces of the 
             proton-coupled electron transfer model as well as the initial 
             nuclear densities $|\chi|^2$ for mass parameters 
             $\mu^{-1} = 100$ and $900$ are shown for a case of weak coupling
             (left, \unit[$R_{\rm c} = 4.0$]{$a_0$}) and strong coupling
             (right, \unit[$R_{\rm c} = 7.0$]{$a_0$}).}
    \label{fig:pes}
  \end{figure}

  To illustrate this continuity equation, we consider a model for proton-coupled 
  electron transfer \cite{shin1995} with the parameters of \cite{eich2016}.
  In this one-dimensional model, two ``ions'' of infinite mass are located at
  $\pm L/2$, and a positively charged particle, the nucleus, as well as a 
  negatively charged particle, the electron, are allowed to move along  
  dimensions $R$ and $r$, respectively.
  The Hamiltonian for the system is
  \begin{align}
    H = -\frac{\mu}{2} \pR^2 + \hat{H}_{\rm el} 
  \end{align}
  where $\mu = m/M$ is the mass ratio between electron and nucleus, and where 
  \begin{align}
    \hat{H}_{\rm el} = -\frac{\pr^2}{2} + \frac{1}{|R - \frac{L}{2}|}
                              + \frac{1}{|R + \frac{L}{2}|}
                              - \frac{\operatorname{erf} \left( \frac{|r-R|}{R_{\rm c}} \right)}{|R - r|}
                              - \frac{\operatorname{erf} \left( \frac{|r-\frac{L}{2}|}{R_{\rm r}} \right)}{|r - \frac{L}{2}|}
                              - \frac{\operatorname{erf} \left( \frac{|r+\frac{L}{2}|}{R_{\rm l}} \right)}{|r + \frac{L}{2}|}.
  \end{align}
  The parameters are taken to be \unit[$L= 19$]{$a_0$}, \unit[$R_{\rm r} = R_{\rm l} = 3.5$]{$a_0$},
  and we consider different values for $R_{\rm c}$.
  This parameter determines the coupling between the two lowest 
  electronic states in a Born-Oppenheimer description of the dynamics.
  The respective Born-Oppenheimer potential energy surfaces are shown in Figure
  \ref{fig:pes} for \unit[$R_{\rm c} = 4.0$]{$a_0$} (weak coupling) and 
  \unit[$R_{\rm c} = 7.0$]{$a_0$} (strong coupling).
  
  As in \cite{eich2016}, we choose the initial state to be 
  $\psi(R,r|0) = G(R-R_0, \sigma) \phi_0^{\rm BO}(r|R)$ with the electronic ground 
  state $\phi_0^{\rm BO}$ within the Born-Oppenheimer approximation and with 
  Gaussian $G$ centered at \unit[$R_0 = 5.0$]{$a_0$} having variance 
  $\sigma^2 = (0.15 \, a_0)^2 \sqrt{2000 \mu}$.
  The specific choice of the parameters for the Gaussian are, however, not 
  important for the following analysis.
  The initial densities for the choices $\mu^{-1} = 100$ and $\mu^{-1} = 900$ 
  are shown in Figure \ref{fig:pes}.
  We integrate the TDSE \eqref{eq:tdse} by constructing a sparse-matrix 
  representation of the Hamiltonian and by applying the corresponding evolution 
  operator to a vector representation of the wavefunction, using the SciPy 
  sparse matrix functionalities \cite{scipy}.
  Videos and further pictures of the dynamics are given in the supplemental 
  material.
  
  We investigate the mass dependence and coupling-strength dependence of the 
  three terms in the electronic continuity equation \eqref{eq:ce8}.
  To quantify the contributions of each term, we integrate their 
  magnitudes over the whole simulation time.
  Subsequently, the results are normalize by a factor $1/\kappa$ such that the 
  maximum value of the largest of these quantities, for the simulation with the 
  largest flux is 1, i.e.\ we define
  \begin{align}
    N_t &= \frac{1}{\kappa} \int\limits_{0}^{t_{\rm max}} \left| \expect{|\chi|^2 \pt |\phi|^2}_R \right| dt \\
%     N_c &= \frac{1}{\kappa} \int\limits_{0}^{t_{\rm max}} \left| \expect{J[\chi,+A] \cdot \pR |\phi|^2}_R \right| dt \\
    N_c &= \frac{1}{\kappa} \int\limits_{0}^{t_{\rm max}} \left| \expect{|\chi|^2 \hat{c} |\phi|^2}_R \right| dt \\
    N_J &= \frac{1}{\kappa} \int\limits_{0}^{t_{\rm max}} \left| \pr \cdot \expect{|\chi|^2 j[\phi]}_R \right| dt.
  \end{align}
  The larger $N_t$, $N_c$, or $N_J$ are, the more flux they correspond to.
  
  \begin{figure}
    \centering
    \includegraphics[width=0.8\textwidth]{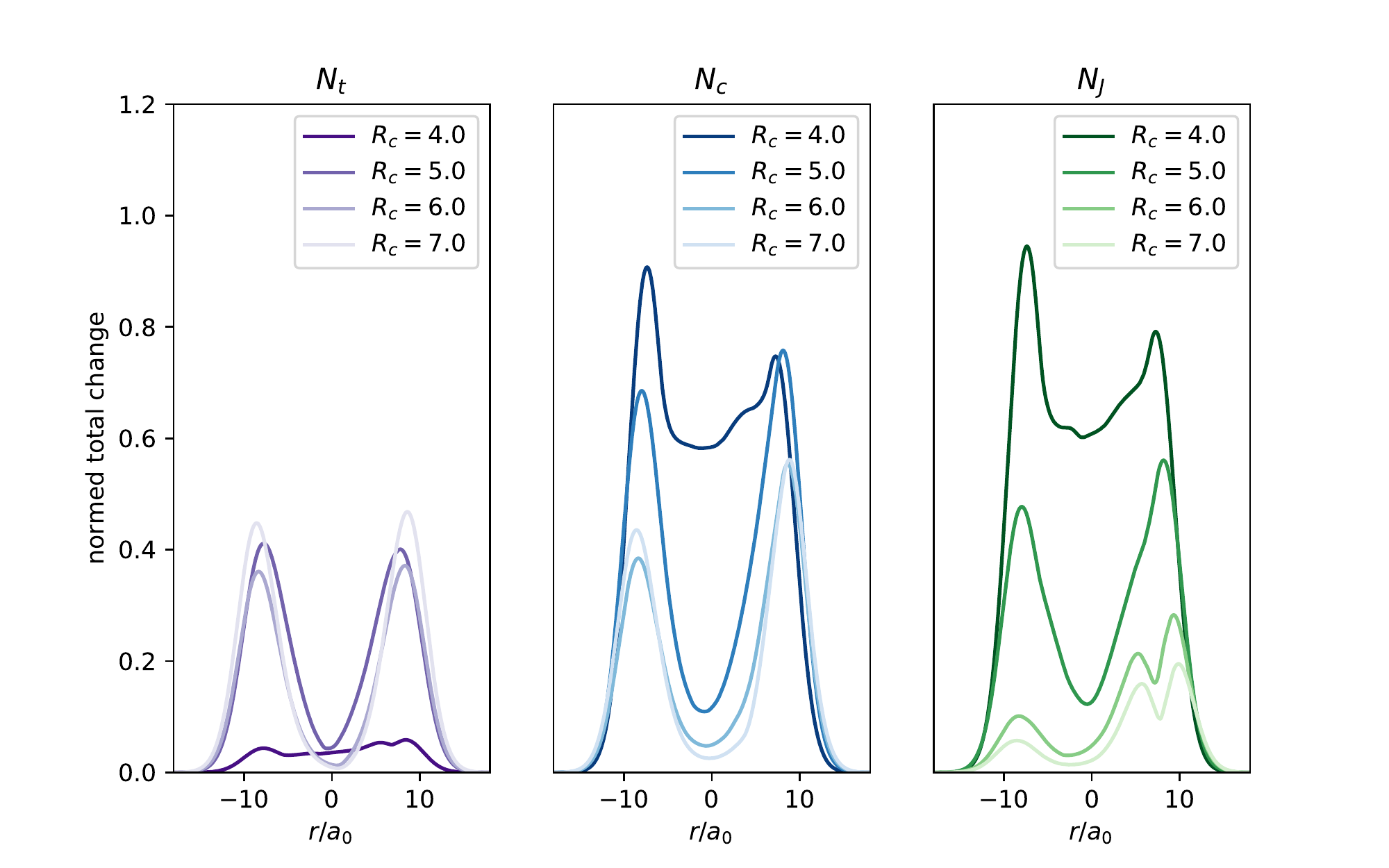}
    \caption{Measures $N_t$, $N_c$, $N_J$ of the three contributions occurring 
             in the continuity equation for the electron density for a mass 
             parameter $\mu^{-1} = 900$, for different coupling parameters $R_{\rm c}$.}
    \label{fig:ttta}
  \end{figure}
  
  The coupling-strength-dependence is shown for $\mu^{-1} = 900$ in Figure 
  \ref{fig:ttta}.
  From the figure, we see that for strong coupling, both the contributions 
  w.r.t.\ the external and the internal clock are important.
  In contrast, for weak coupling $N_t$ becomes negligible compared to the 
  other two terms, i.e., the external clock plays a minor role.
  
  \begin{figure}
    \centering
    \includegraphics[width=0.8\textwidth]{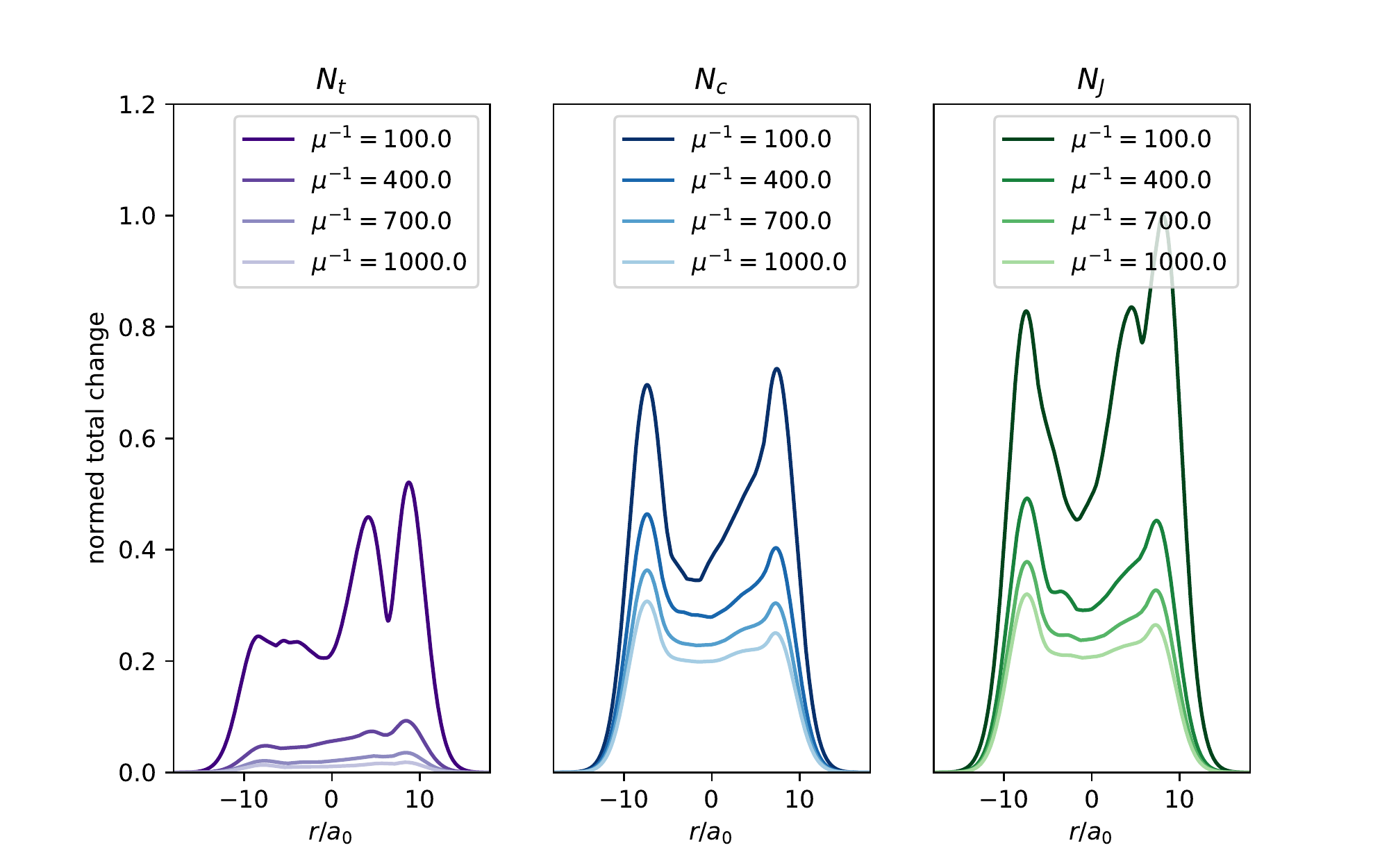}
    \caption{Measures $N_t$, $N_c$, $N_J$ of the three contributions occurring 
             in the continuity equation for the electron density for a coupling
             parameter \unit[$R_{\rm c} = 4.0$]{$a_0$}, for different mass parameters
             $\mu^{-1}$.}
    \label{fig:tttb}
  \end{figure}
  
  The other relevant factor for the dynamics is the mass ratio.
  The mass-dependence of $N_t, N_c, N_J$ for weak coupling
  (\unit[$R_{\rm c} = 4$]{$a_0$}) is shown in Figure \ref{fig:tttb}.
  We see that $N_t$ becomes less and less important with increasing mass.
  This is a general trend, i.e., for weak couplings and large masses the 
  contribution of the external clock can be neglected, whereas for small masses 
  and/or strong couplings all terms are relevant.
  
  The conditions where only the nucleus as internal clock is relevant
  for the electron dynamics (large mass ratio, small couling parameter) are 
  also those for which the Born-Oppenheimer approximation is valid.
  There, the molecular wavefunction is written as
  \begin{align}
    \psi(R,r|t) \approx \chi^{\rm BO}(R|t) \phi^{\rm BO}(r|R) \label{eq:psi_bo}
  \end{align}
  with $\phi^{\rm BO}(r|R)$ being an eigenfunction of $\hat{H}_{\rm el}$ at a fixed nuclear 
  configuration $R$.
  It can be derived from the Exact Factorization in the limit of vanishing 
  mass ratio $\mu$ of electronic and nuclear mass, cf.\ \cite{eich2016}, 
  which leads to the disappearance of $i \hbar \pt$, $\hat{U}$, and 
  $\hat{C}$ in the conditional equation \eqref{eq:phi_t}.
  
  The Born-Oppenheimer approximation is usually interpreted such that due to 
  their comparably small mass, the electrons react instantaneously to a change 
  in the nuclear position.
  In this way, the electron density is well approximated by the density of an
  eigenstate $\phi^{\rm BO}$ of $\hat{H}_{\rm el}$ at a given nuclear configuration $R$.
  We see the meaning of this interpretation in the vanishing of the 
  contribution $\pt |\phi|^2$ to the continuity equation, i.e., in the 
  vanishing of the (direct) dependence of the electronic density on the 
  external time.
  However, there is still a motion of the electrons which is given by the 
  electronic flux density $j[\phi]$, but this motion is induced by the motion 
  of the nuclei.
  In the continuity equation, this part corresponds to the clock-dependent 
  contribution $\hat{c} |\phi|^2$ (that indirectly depends on $t$, too), which 
  entirely cancels the divergence of the electronic flux density for large 
  nuclear masses and small coupling strengths.
  
  Typically, the Born-Oppenheimer approximation does not take into account 
  this electronic motion w.r.t.\ the internal clock of the nuclei.
  The consequence is that the electronic density 
  \begin{align}
    \rho_{\rm el} := \int |\psi|^2 dR \approx \int |\chi^{\rm BO}(R|t)|^2 |\phi^{\rm BO}(r|R)|^2 dR,
    \label{eq:rho_bo}
  \end{align}
  is close to the true density, but the flux density computed with the 
  Born-Oppenheimer wavefunction is zero \cite{barth2009},
  \begin{align}
    j_{\rm el} = \frac{\hbar}{m_e} \int \operatorname{Im} \left( \cpsi \pr \psi \right) 
           \approx 
           \frac{\hbar}{m_e} \int |\chi^{\rm BO}|^2 \operatorname{Im} \left( \cphi^{\rm BO} \pr \phi^{\rm BO} \right) 
           \equiv 0.
    \label{eq:j_bo}
  \end{align}
  Comparing \eqref{eq:rho_bo} and \eqref{eq:j_bo}, we see that the continuity 
  equation \eqref{eq:eoc_gen_dif} seems to be invalid.
  A number of investigations \cite{diestler2013,diestler2013b,scherrer2013,schild2016}
  partly clarify the issue.
  In view of the results presented above, the failure of the Born-Oppenheimer 
  wavefunction can be interpreted as the lack of correctly accounting for the
  motion w.r.t.\ the internal clock.
  This interpretation is in line with \cite{schild2016}, where it was shown that 
  the electronic flux density can be recovered for conditions where the 
  Born-Oppenheimer approximation is valid if $\hat{C}$ is treated as 
  perturbation, i.e., if the motion is referred to the nuclei being the clock. 
  
  We note that a different way to find an approximation for the motion of the 
  electrons relative to the nuclear clock is to replace the CDSE with a TDSE.
  This can be done by starting from the conditional equation \eqref{eq:phi}, 
  applied to the electron-nuclear problem (i.e., without dependence on the 
  external time $t$), and taking the classical limit for the nuclear 
  wavefunction.
  The result are essentially the equations of motion of Ehrenfest molecular 
  dynamics \cite{tully1998}.
  We find that a simulation of the electron dynamics using equations of this 
  type, specifically solving
  \begin{align}
    i \hbar \partial_T \phi(r|T) = \hat{H}_{\rm el}(T) \phi(r|T). \label{eq:ehrlike}
  \end{align}
  for effective time $T$ (which is close to the external time $t$) defined via 
  the expectation values of nuclear position and momentum, 
  $T := \expect{R(t)} / \langle\hat{P}(t) \rangle$, yields a good approximation for 
  the electronic flux density in our models.
  In the classical limit of the nuclei, the internal classical time $T$ and the 
  external time $t$ are of course identical.
  We note that the electronic Hamiltonian $\hat{H}_{\rm el}$ depends on $T$ in the 
  sense that its dependence on the nuclear position $R$ is replaced by the 
  expectation value $\expect{R}$.
  
  {\bf Conclusion and outlook}
  
  In this article, we start with the premise that time is obtained in the 
  classical limit of a clock and that the system depends conditionally on the 
  configuration of the clock.
  From a quantum-mechanical perspective, it follows that there is no time,
  there are only clocks.
  We investigate the consequences of this statement on the continuity equation,
  finding that the flux density of the clock plays a vital role if the clock
  needs to be treated as a quantum system.
  
  The generalization of the TDSE to a CDSE by means of the Exact Factorization
  allows to define any degree of freedom as a clock.
  Interpreting part of a quantum system as clock for the remaining degrees of 
  freedom can be a helpful idea for developing effective simulation methods.
  The clock-dependent continuity equation \eqref{eq:cc} or its analogue for 
  dynamics referred to an external time, \eqref{eq:ccc_t}, may be used as a 
  tool to analyze a general quantum dynamics for possible degrees of freedom 
  that are amenable to such approximate treatments.
  Then, one may use a Born-Oppenheimer-like approach based on the separation of 
  time scales, or one may derive other approximations to the CDSE based on 
  the quasi-classical behavior of some degrees of freedom, like Ehrenfest 
  molecular dynamics, where the CDSE is replaced by a TDSE.
  Our analysis of a simple model of a coupled electron-nuclear dynamics 
  illustrates the idea of a quantum clock and shows in a novel way how the
  Born-Oppenheimer approach works, and might provide some ideas for a time-scale 
  separation of other problems.
  
  Finally, an important point that we discussed only briefly here is the 
  origin of the dynamics, i.e., of evolution without time as a fundamental 
  variable.
  Starting from a stationary state of the universe without reference to an 
  external clock or observer, an internally consistent static theory may be 
  derived.
  However, the mechanism of how a change of the system happens needs to be 
  explored further.
  One way that this problem might conceptually be solved is by inclusion of an 
  internal observer which measures the clock and the properties of the system.
  In this author's opinion, there is some promising on this topic but there is 
  still no relative point of view on time (and space) that yields a fully 
  developed picture of dynamics without time.
  Thus, further work is needed.
  
  {\bf Acknowledgement}
  
  The author is grateful to Basile F.\ E.\ Curchod for helpful comments on 
  the manuscript and the Swiss National Science Foundation for funding this 
  research.
  
  \bibliography{bib}{}
  \bibliographystyle{unsrt}
  
\end{document}